\documentclass[11pt]{article}
\usepackage{amsfonts}
\usepackage{graphicx}

\begin{document}
\begin{center}
\Large{{\sf Renormalization-Group Improvement of Effective Actions Beyond 
Summation  of Leading Logarithms}}

\vskip .8cm

\large{M. R. Ahmady,$^{a}$ V. Elias,$^{b,c}$ D. G. C. McKeon, 
$^{c}$\\ A. Squires,$^{a}$ and T. G. Steele$^{d}$}

\vskip .8cm
{\small \it
$^a$ 
Department of Physics,\\ Mount Allison University, Sackville, NB  E4L 1E6,
Canada
\\
\smallskip
$^b$ 
Perimeter Institute for Theoretical Physics,\\ 35 King Street North,
Waterloo, Ontario  N2J 2W9,  Canada
\\
\smallskip
$^c$ 
Department of Applied Mathematics,\\ The University of Western Ontario,
London, Ontario  N6A 5B7, Canada
\\
\smallskip
$^d$ 
Department of Physics and Engineering Physics,\\ University of Saskatchewan,
Saskatoon, Saskatchewan  S7N 5E2, Canada
}
\end{center}

\vskip 1cm

\abstract{Invariance of the effective action under changes of the renormalization scale 
$\mu$ leads to relations between those (presumably calculated) terms
independent of $\mu$ at a given order of perturbation theory and those higher
order terms dependent on logarithms of $\mu$.  This relationship leads to
differential equations for a sequence of functions, the solutions of which
give closed form expressions for the sum of all leading logs, next to leading
logs and subsequent subleading logarithmic contributions to the effective
action.  The renormalization group is thus shown to provide
information about a model beyond the scale dependence of the model's couplings
and masses.  This procedure is illustrated using
the $\phi_6^3$ model and Yang-Mills theory.  In the latter instance, it
is also shown by using a modified summation procedure that the $\mu$
dependence of the effective action resides solely in a multiplicative
factor of $g^2 (\mu)$ (the running coupling).  This approach is also
shown to lead to a novel expansion for the running coupling in terms of
the one-loop coupling that does not require an order-by-order
redefinition of the scale factor $\Lambda_{QCD}$.  Finally, logarithmic
contributions of the instanton size to the effective action of an SU(2)
gauge theory are summed, allowing a determination of the asymptotic dependence
on the instanton size $\rho$ as $\rho$ goes to infinity to all orders in the
SU(2) coupling constant.}

\section{Introduction}

It is well understood that self-interactions in quantum field theory can serve to 
rescale (renormalize) the parameters that characterize the theory.  Indeed, 
renormalization is usually viewed as necessary to eliminate divergences which 
arise when physical processes are computed.  In the course of this renormalization, 
a mass scale $\mu$ inevitably arises, apparently rendering the results of any 
computation inherently ambiguous.   The effect is illusory [1,2]; one can 
compensate for changes in $\mu$ by concomitant changes in the couplings, masses and 
field strengths characterizing the theory under consideration.  Indeed, it is 
this observation, which ultimately implies that all physical quantities in the 
theory should be independent of $\mu$, which leads to the renormalization-group 
(RG) equation.  When calculating quantities to only a finite order in perturbation
theory, however, there remains a residual dependence on $\mu$.

The formal solution of the RG equation results in a dependence of the couplings 
and masses on the parameter $\mu$ in a manner fixed by the so-called RG functions.  
Generally speaking, these functions are determined by the relationship between 
the renormalized and (presumably infinite) bare couplings in the context of some 
regulating scheme, such as dimensional regularization [3]. However, when one uses 
$\zeta$-function regularization [4] or its multiloop generalization, operator 
regularization [5], no explicit divergences appear when the regulating parameter 
vanishes; explicit renormalization is not required to excise infinities in these 
schemes.  A mass scale $\mu$ nevertheless arises in the course of using such schemes. 
Hence, in this approach one is forced to determine the RG functions by direct 
examination of the $\mu$-dependence of the effective action. This approach can in
principle also be used to determine the RG functions when one is using dimensional
regularization, though in practice the RG functions  in dimensional regularization 
are extracted by considering the relationship between bare and renormalized
quantities. Such an approach has 
been explicitly developed to two-loop order in two different massive scalar field 
theories, one with a cubic self-interaction coupling in six dimensions [6], 
and one with the usual quartic self-interaction coupling in four dimensions [7].
        
However, the RG equation can do more than just provide the usual mass scale dependence 
to the couplings and masses in a given model.  Consistency conditions arise which 
relate logarithms occurring in higher loop calculations to non-logarithmic contributions 
that occur at lower orders.  In dimensional regularization, these correspond 
to relationships between coefficients of poles appearing in the equation 
relating the bare and renormalized couplings [2,8].  Such consistency conditions also 
constitute relations between the coefficients of logarithms appearing in the effective 
action [9], in both operator regularization and dimensional regularization. 
        
In the present paper, we exploit these relationships to ``RG-improve'' such effective 
actions via explicit summation of leading and successively subleading logarithms 
accessible from the RG-equation to {\it all} orders in perturbation theory. Such an approach has already been applied to a number of  
$\overline{MS}$ perturbative processes [10] as well as to the relationship between bare and 
renormalized coupling constants in the context of dimensional regularization [8]. 
In Section 2, we continue the work of ref.\ [9], in which leading logarithms were 
summed to all orders, by summing to all orders in perturbation theory the 
next-to-leading-logarithmic radiative corrections to the classical action for a 
massive scalar field theory in six dimensions with trilinear self-interaction 
coupling $(\phi_6^3)$.  This simple field theoretical example is particularly useful in that 
the effective action involves separate kinetic, mass and self-interaction radiative 
corrections. The procedures delineated in Section 2 can easily be extended to 
all-orders summation of subsequent nonleading sets of logarithms, once appropriate 
perturbative calculations beyond those of ref.\ [6] become available.  The procedure
used is related to one employed in \cite{11,12} in the context of the effective potential
for a massive $\phi_4^4$ model, as well as that employed in ref.\ [13] to sum the
leading-logarithm contributions to the effective action of QED.

In Section 3, we use an approach first delineated in the Appendix to ref.\ [10] to find an all-orders solution to the RG-equation for the kinetic 
term of the $\phi_6^3$ theory via a restructured version of the series for this kinetic term 
obtained in Section 2. The consistency of this closed-form expression for the kinetic term with the 
series expression extracted in Section 2 is demonstrated in Appendix A.

In Section 4 we first apply the methods of Section 2 to obtain the all-orders 
summation of leading and the first three subsequent subleading sets of logarithmic 
corrections to the effective action for Yang-Mills theory.  We then apply the 
methods developed in Section 3 to an appropriately restructured series for this 
effective action. For this latter approach, we are able to determine in full the 
coefficient of $F_{\mu\nu a} F^{\mu\nu a}$  if the $\beta$-function is known to 
all orders and the coefficient is known for a particular value of $\mu$. Such
an approach is particularly relevant for $N = 1$ supersymmetric 
Yang-Mills theory, for which the $\beta$-function is entirely known [14,15]. 
Moreover, we are also able to demonstrate that all the $\mu$-dependence of this
coefficient is necessarily proportional to the running gauge coupling constant $g^2(\mu)$. 

In Section 5 we demonstrate how a direct comparison of the two approaches 
delineated in Section 4 can be employed to extract an alternate series solution for the 
running coupling constant. As a check on our approach, a derivation of this same 
series based upon explicit use of RG invariance is presented in Appendix B. For 
SU(N) vector-like gauge theories, such as QCD, the series we obtain enables us to 
express the four-loop running couplant as a series in which the na\"{i}ve one-loop 
running couplant surprisingly appears as the expansion  parameter. We conclude Section 5 by 
discussing in detail the departure of this series in the infrared region from the 
true running couplant's behaviour, as determined from explicit integration over the couplant's 
$\beta$-function. Such a departure is shown to reflect the disparity between the Landau 
singularity in the one-loop couplant, which is the series expansion parameter, and 
the Landau singularity characterising evolution via the full four-loop 
order $\beta$-function.

Finally, in Section 6 we consider how the instanton contribution to the effective
Lagrangian of an SU(2) gauge theory is affected by all-orders summation of logarithms 
of the instanton size. Since the instanton size $\rho$ only appears in logarithmic
contributions to the perturbative series in the SU(2) gauge coupling constant,
it is possible to sum arbitrarily-subleading logarithms in order to obtain asymptotic
dependence on $\rho$ (as $\rho$ becomes large) to {\it all} orders in the coupling
constant.  We find that the asymptotic behaviour of the integral over instanton 
size retains essentially the same infrared divergence previously extracted by 
't Hooft \cite{16} from the one-loop contribution, with only a slight logarithmic 
modification arising from the summation of next-to-leading logs. The summation of subsequent
logarithms in the perturbative series ({\it i.e.} logarithms other than leading and next-to-leading) are shown not to contribute to the asymptotic $\rho$-dependence in the large-$\rho$ limit.

\newpage


\renewcommand{\theequation}{2.\arabic{equation}}
\setcounter{equation}{0}

\section{RG-Resummation of the $\phi_6^3$ Effective Action}

In ref.\ \cite{6}, the leading logarithms within the effective action for a
scalar field ($f$) theory with a trilinear self-coupling in six dimensions are
computed.  These leading logarithms are summed in ref.\ [9].  The classical action for 
such a theory in Euclidean space is just
\begin{equation}
\Gamma_0 \left[ f(x) \right] = - \frac{1}{2} f(x) \partial^2 f(x) + \frac{m^2}{2} f^2 (x) + \frac{\lambda f^3 (x)}{3!}.
\end{equation}
Radiative corrections to eq.\ (2.1) leave us with an effective action of the
form
\begin{equation}
\Gamma [f] = - \frac{1}{2} f \partial^2 f A (\lambda, L) + \frac{m^2}{2} f^2 B(\lambda, L) + \frac{\lambda f^3}{3!} C(\lambda, L)
\end{equation}
with $L \equiv \log\left(m^2/\mu^2\right)$ and with
\begin{equation}
A(\lambda, L) = \sum_{n=0}^\infty \sum_{m=n}^\infty a_{m,n} \lambda^{2m} L^n,
\end{equation}
\begin{equation}
B(\lambda, L) = \sum_{n=0}^\infty \sum_{m=n}^\infty b_{m,n} \lambda^{2m} L^n,
\end{equation}
\begin{equation}
C(\lambda, L) = \sum_{n=0}^\infty \sum_{m=n}^\infty c_{m,n} \lambda^{2m} L^n.
\end{equation}
Since the effective action (2.2) must be independent of $\mu$, we have
the RGE
\begin{equation}
\mu \frac{d\Gamma}{d\mu} = 0 = \left( \mu \frac{\partial}{\partial\mu} + \beta (\lambda) \frac{\partial}{\partial\lambda} 
- \gamma_m (\lambda) m^2 \frac{\partial}{\partial m^2} - \gamma_\Gamma
(\lambda) \int d^4 x' f(x') \frac{\delta}{\delta f(x')} \right) \Gamma,
\end{equation}
where the RG functions within eq.\ (2.6) are defined as follows:
\begin{equation}
\beta(\lambda) \equiv \mu \frac{d\lambda}{d\mu} = \sum_{n=1}^\infty
B_{2n+1} \lambda^{2n+1},
\end{equation}
\begin{equation}
-\gamma_m (\lambda) m^2 \equiv \mu \frac{dm^2}{d\mu} = - m^2
\sum_{n=1}^\infty G_{2n} \lambda^{2n},
\end{equation}
\begin{equation}
-\gamma_\Gamma (\lambda) f \equiv \mu \frac{df}{d\mu} = -f \sum_{n=1}^\infty
D_{2n} \lambda^{2n}.
\end{equation}
Noting for an arbitrary function $F(\lambda, L)$ that $\mu \frac{\partial}{\partial\mu} F(\lambda , L) = -
2\frac{\partial}{\partial L} F(\lambda , L)$ and that
$m^2 \frac{\partial}{\partial m^2} F(\lambda , L) = 
\frac{\partial}{\partial L} F(\lambda , L)$, we substitute eq.\ (2.2)
into eq.\ (2.6) to obtain the following RG-equations for $A(\lambda , L)$,
$B(\lambda , L)$ and $C(\lambda , L)$:
\begin{equation}
\left[ -\left( 2+\gamma_m (\lambda) \right) \frac{\partial}{\partial L}
+ \beta(\lambda) \frac{\partial}{\partial\lambda} - 2\gamma_\Gamma
(\lambda) \right] A(\lambda, L) = 0,
\end{equation}
\begin{equation}
\left[ -\left( 2+\gamma_m (\lambda) \right) \frac{\partial}{\partial L}
+ \beta(\lambda) \frac{\partial}{\partial\lambda} - \left( 2\gamma_\Gamma
(\lambda) + \gamma_m (\lambda) \right) \right] B(\lambda, L) = 0,
\end{equation}
\begin{equation}
\left[ -\left( 2+\gamma_m (\lambda) \right) \frac{\partial}{\partial L}
+ \beta(\lambda) \frac{\partial}{\partial\lambda} + \left( \frac{\beta(\lambda)}{\lambda} - 3\gamma_\Gamma (\lambda) \right)
\right] C(\lambda, L) = 0.
\end{equation}
Upon substitution of eq.\ (2.3) and eqs.\ (2.7-2.9) into eq.\ (2.10), we find the
aggregate coefficients of $\lambda^{2n} L^{n-1}$ and $\lambda^{2n}
L^{n-2}$ on the left hand side of eq.\ (2.10) respectively vanish provided the
following recursion relations are upheld:
\begin{equation}
-2n a_{n,n} + \left[ 2(n-1) B_3 - 2D_2 \right] a_{n-1, n-1} = 0,
\end{equation}
\begin{eqnarray}
&& -2 (n-1) a_{n, n-1} + \left[ 2(n-1) B_3 - 2D_2 \right] a_{n-1, n-2}
\nonumber\\
&& -(n-1) G_2 a_{n-1, n-1} + \left[ 2(n-2) B_5 - 2D_4 \right] a_{n-2,
n-2} = 0.
\end{eqnarray}
Substitution of eq.\ (2.4) and eqs.\ (2.7-2.9) into eq.\ (2.11) also yields
recursion relations for $b_{n,m}$:
\begin{equation}
-2n b_{n,n} + \left[ 2(n-1) B_3 - 2D_2 - G_2 \right] b_{n-1, n-1} = 0,
\end{equation}
\begin{eqnarray}
-2(n-1) b_{n, n-1} + \left[ 2(n-1) B_3 - 2D_2 - G_2 \right] b_{n-1, n-
2}\nonumber\\
-(n-1) G_2 b_{n-1, n-1} + \left[ 2(n-2) B_5 - 2D_4 - G_4 \right] b_{n-2,
n-2} = 0,
\end{eqnarray}
and similar substitution of eq.\ (2.5) and eqs.\ (2.7-2.9) into eq.\ (2.12) yields
the following recursion relations for $c_{n,m}$:
\begin{equation}
-2n c_{n,n} + \left[ (2n-1) B_3 - 3D_2 \right] c_{n-1, n-1} = 0,
\end{equation}
\begin{eqnarray}
-2(n-1) c_{n,n-1} + \left[ (2n-1) B_3 - 3 D_2 \right] c_{n-1, n-
2}\nonumber\\
-(n-1) G_2 c_{n-1, n-1} + \left[ (2n-3) B_5 - 3D_4 \right] c_{n-2, n-2}
= 0.
\end{eqnarray}
An explicit perturbative computation \cite{7} produces the following values
for the leading two orders of coefficients in (2.3-2.5) $[k \equiv
1/(4\pi)^3]$:
\begin{equation}
a_{0,0} = b_{0,0} = c_{0,0} \equiv 1,
\end{equation}
\begin{equation}
a_{1,0} = 0, \; \; \; b_{1,0} = k/2, \; \; \; c_{1,0} = 0,
\end{equation}
\begin{equation}
a_{1,1} = -k/12, \; \; \; b_{1,1} = -k/2, \; \; \; c_{1,1} = -k/2,
\end{equation}
\begin{equation}
a_{2,1} = k^2/216, \; \; \; b_{2,1} = -43k^2/48, \; \; \; c_{2,1} = -
13k^2/48,
\end{equation}
\begin{equation}
a_{2,2} = 5k^2/144, \; \; \; b_{2,2} = 5k^2/16, \; \; \; c_{2,2} = 5k^2/16.
\end{equation}
If we choose $n = 1$ in eqs.\ (2.13), (2.15) and (2.17), we  find from the 
coefficients (2.19) and (2.21)
that the lead terms of the RG-functions (2.7-2.9) are
\begin{equation}
B_3 = -3k/4, \; \; \; G_2 = 5k/6, \; \; \; D_2 = k/12.
\end{equation}
Indeed, these results are sufficient to determine the coefficients (2.23) by setting 
$n = 2$ in eqs.\ (2.13), (2.15) and (2.17). Similarly one can substitute coefficients (2.19-2.22) 
and leading terms (2.24) into the $n = 2$ versions of the recursion relations (2.14), (2.16) and (2.18) to obtain
the next-to-leading contributions to the RG functions (2.7-2.9):
\begin{equation}
B_5 = -125 k^2/144, \; \; \; G_4 = 97 k^2/108, \; \; \; D_4 = 13k^2/432.
\end{equation}
The RG-function results (2.24) and (2.25) plus the recursion relations (2.13-18) permit one to sum the leading
and next-to-leading logarithmic contributions to $A$, $B$, and $C$ to {\it all}
orders of perturbation theory. The leading-logarithm (LL) contribution to $A$, as given by
(2.3), is just
\begin{eqnarray}
\left( A(\lambda, L)\right)_{LL} & = & a_{0,0} + a_{1,1} \lambda^2 L + a_{2,2} (\lambda^2 L)^2 + a_{3,3} (\lambda^2 L)^3 + ...\nonumber\\
& = & \sum_{n=0}^\infty a_{n,n} (\lambda^2 L)^n \equiv R_0 (\lambda^2 L).
\end{eqnarray}
If we define $u \equiv \lambda^2L$ and multiply the recursion
relation (2.13) by $u^{n-1}$ and then sum over $n$ from $n = 1$
to $\infty$, we obtain the differential equation
\begin{equation}
(1-B_3 u) R_0^{\prime} (u) + D_2 R_0 (u) = 0
\end{equation}
with initial condition $R_0(0) = a_{0,0} = 1$. The solution to this
differential equation is
\begin{equation}
R_0 (\lambda^2 L) = (1 - B_3 \lambda^2 L)^{D_2/B_3} = (1 + 3\lambda^2 L / 256\pi^3)^{-1/9}.
\end{equation}
To find $LL$ contributions to $B$ and $C$,
\begin{equation}
\left( B(\lambda, L)\right)_{LL} = \sum_{n=0}^\infty b_{n,n} (\lambda^2 L)^n 
\equiv S_0 (\lambda^2 L),
\end{equation}
\begin{equation}
\left( C(\lambda, L)\right)_{LL} = \sum_{n=0}^\infty c_{n,n} (\lambda^2 L)^n 
\equiv T_0 (\lambda^2 L),
\end{equation}
one may apply exactly the same procedure as above to obtain differential
equations for $S_0$ and $T_0$ from the recursion relations (2.15) and 
(2.17):
\begin{equation}
(1 - B_3 u)  S_0^{\prime} (u) + (D_2 + G_2 / 2) S_0 (u) = 0,
\end{equation}
\begin{equation}
(1 - B_3 u) T_0^{\prime} + (3D_2 / 2 - B_3 / 2) T_0 (u) = 0.
\end{equation}
Since $S_0(0) = b_{0,0} = 1$ and $T_0(0) = c_{0,0} = 1$, we easily find
the solutions
\begin{equation}
S_0 (\lambda^2 L) = (1 - B_3 \lambda^2 L)^{(D_2 + \frac{1}{2} G_2)/B_3}
= (1 + 3 \lambda^2 L / 256\pi^3)^{-2/3},
\end{equation}
\begin{equation}
T_0 (\lambda^2 L) = (1 - B_3 \lambda^2 L)^{(3D_2 - B_3)/2B_3}
= (1 + 3 \lambda^2 L / 256\pi^3)^{-2/3}.
\end{equation}
It is curious that the summations of LL contributions to $B$ and $C$
are identical.

The recursion relations (2.14), (2.16), and (2.18) may be similarly employed
to determine the next-to-leading logarithm (NLL) contributions to
$A(\lambda , L)$, $B(\lambda , L)$ and $C(\lambda , L)$, as given by eqs.\ (2.3-2.5):
\begin{eqnarray}
\left[ A(\lambda, L) \right]_{NLL} & = & a_{1,0} \lambda^2 + a_{2,1}
\lambda^4 L + a_{3,2} \lambda^6 L^2 + ...\nonumber\\
& = & \lambda^2 \sum_{n=1}^\infty a_{n, n-1} (\lambda^2 L)^{n-1} \equiv
\lambda ^2 R_1 (\lambda^2 L),
\end{eqnarray}
\begin{equation}
\left[ B(\lambda, L)\right]_{NLL} = \lambda^2 \sum_{n=1}^\infty b_{n, n-
1} (\lambda^2 L)^{n-1} \equiv \lambda^2 S_1 (\lambda^2 L),
\end{equation}
\begin{equation}
\left[ C(\lambda, L)\right]_{NLL} = \lambda^2 \sum_{n=1}^\infty c_{n, n-
1} (\lambda^2 L)^{n-1} \equiv \lambda^2 T_1 (\lambda^2 L).
\end{equation}
Upon multiplication of the recursion relations (2.14), (2.16) and (2.18) by
$u^{n-2}$ followed by summation from $n = 2$ to $\infty$, the series
definitions (2.26), (2.29), and (2.30) for $R_0(u)$, $S_0(u)$, and $T_0(u)$,
and the definitions (2.35-2.37) for $R_1(u)$, $S_1(u)$, and $T_1(u)$
lead to the following linear first order differential equations for the
latter three quantities:
\begin{equation}
(1-B_3 u) \frac{dR_1(u)}{du} + (D_2 - B_3) R_1 (u) = \left[ \left( B_5 u - \frac{G_2}{2}\right) \frac{d}{du} - D_4 \right] R_0 (u),
\end{equation}
\begin{equation}
(1 - B_3 u) \frac{d S_1 (u)}{du} + \left( D_2 + \frac{G_2}{2} - B_3 \right) S_1 (u) = \left[ \left( B_5 u - \frac{G_2}{2} \right) \frac{d}{du} - D_4 - \frac{G_4}{2} \right] S_0 (u),
\end{equation}
\begin{equation}
(1 - B_3 u) \frac{d T_1 (u)}{du} + \left( \frac{3D_2}{2} - B_3 \right) T_1 (u) = \left[ \left( B_5 u - \frac{G_2}{2} \right) \frac{d}{du} - \frac{3D_4}{2} + \frac{B_5}{2} \right] T_0 (u).
\end{equation}
Given the explicit solutions (2.28), (2.33), (2.34) already obtained for $R_0$,
$S_0$, and $T_0$, and given the initial conditions $R_1(0) = a_{1,0} =
0$, $S_1(0) = b_{1,0} = k/2$, $T_{1,0} = c_{1,0} = 0$, we obtain the
following solutions to eqs.\ (2.38-2.40):
\begin{eqnarray}
R_1(u) & = & \left[ \frac{D_4}{B_3} - \frac{D_2 B_5}{B_3^2} \right] \left( 1 - B_3 u \right)^{D_2/B_3}\nonumber\\
& + & \frac{\left[ a_{1,0} - \left[ \frac{D_4}{B_3} - \frac{D_2 B_5}{B_3^2} \right] - \frac{D_2}{B_3} \left( \frac{G_2}{2} - \frac{B_5}{B_3} \right) \log \left( 1 - B_3 u \right) \right] }{\left( 1 - B_3 u \right)^{1 - D_2/B_3}}\nonumber\\
& = & \frac{1}{(4\pi)^3} \left\{ \frac{43}{486} \left( 1 + \frac{3u}{256\pi^3} \right)^{-1/9} - \left[ \frac{43}{486} + \frac{20}{243} \log \left( 1 + \frac{3u}{256\pi^3} \right) \right] \left( 1 + \frac{3u}{256\pi^3} \right)^{-10/9} \right\},\nonumber\\
\end{eqnarray}
\begin{eqnarray}
S_1 (u) & = & \left[ -\frac{(2D_2 + G_2)B_5}{2B_3^2} + \frac{(2D_4 + G_4)}{2B_3} \right] \left( 1 - B_3 u \right) ^{\frac{2D_2 + G_2}{2B_3}}\nonumber\\
& + & \left\{ b_{1,0} - \left[ \frac{(2D_4 + G_4)}{2B_3} - \frac{(2D_2 + G_2)}{2B_3^2} B_5 \right] \right.\nonumber\\
& - & \left. \frac{(2D_2 + G_2)}{2B_3} \left( \frac{G_2}{2} - \frac{B_5}{B_3} \right) \log \left( 1 - B_3 u \right) \right\} \left( 1 - B_3 u \right)^{\frac{2D_2 + G_2 - 2B_3}{2B_3}}\nonumber\\
& = & \frac{1}{(4\pi)^3} \left\{ \frac{43}{324} \left( 1 + \frac{3u}{256\pi^3} \right)^{-2/3} + \left[ \frac{119}{324} - \frac{40}{81} \log \left( 1 + \frac{3u}{256\pi^3} \right) \right] \left( 1 + \frac{3u}{256\pi^3} \right)^{-5/3} \right\},\nonumber\\
\end{eqnarray}
\begin{eqnarray}
T_1(u) & = & \left[ -\frac{(3D_2 - B_3)}{2B_3^2} B_5 + \frac{(3D_4 - B_5)}{2B_3} \right] \left( 1 - B_3 u \right)^{\frac{3D_2 - B_3}{2B_3}}\nonumber\\
& + & \left\{ c_{1,0} - \left[ \frac{(3D_4 - B_5)}{2 B_3} - \frac{(3D_2 - B_3)B_5}{2 B_3^2} \right] \right.\nonumber\\
& - & \left. \frac{(3D_2 - B_3)}{2B_3} \left( \frac{G_2}{2} - \frac{B_5}{B_3} \right) \log \left( 1 - B_3 u \right) \right\} \left( 1 - B_3 u \right)^{\frac{3(D_2 - B_3)}{2 B_3}}\nonumber\\
& = & \frac{1}{(4\pi)^3} \left\{ \frac{43}{324} \left( 1 + \frac{3u}{256\pi^3} \right)^{-2/3} - \left[ \frac{43}{324} + \frac{40}{81} \log \left( 1 + \frac{3u}{256\pi^3} \right) \right] \left( 1 + \frac{3u}{256\pi^3} \right)^{-5/3} \right\}.\nonumber\\
\end{eqnarray}
For the procedures delineated above to be extended to the summation of
subsequent subleading logs, one must have perturbatively-calculated
values
of the coefficients $a_{k,0}$, $b_{k,0}$, $c_{k,0}$ with $k \geq 3$. One
is then able to utilize subsequent recursion relations devolving from
the $RGE$'s (2.10-12) to obtain first-order differential equations for
\begin{equation}
\left( \begin{array}{cc} R_k (u)\\S_k (u)\\T_k (u)\end{array} \right) = \sum_{n=k}^\infty
\left( \begin{array}{cc} a_{n, n-k}\\b_{n, n-k}\\c_{n, n-k} \end{array}\right) u^{n-k}
\end{equation}
with known initial values $R_k(0) = a_{k,0}$, $S_k(0) = b_{k,0}$,
$T_k(0) = c_{k,0}$. Solutions of these differential equations for progressively larger 
values of $k$ allows one to calculate $A$, $B$ and $C$ as a succession of all-orders 
summations of successively-subleading logarithms
\begin{equation}
A(\lambda, L) = \left( A(\lambda, L) \right)_{LL} + \left( A(\lambda, L) \right)_{NLL} + ... = \sum_{k=0}^\infty \lambda^{2k} R_k (\lambda^2 L),
\end{equation}
\begin{equation}
B(\lambda, L) = \sum_{k=0}^\infty \lambda^{2k} S_k (\lambda^2 L),
\end{equation}
\begin{equation}
C(\lambda, L) = \sum_{k=o}^\infty \lambda^{2k} T_k (\lambda^2 L).
\end{equation}

\newpage


\renewcommand{\theequation}{3.\arabic{equation}}
\setcounter{equation}{0}

\section{Closed-Form All-Orders Solution to $A(\lambda, L)$}

An alternative formulation of the summation of logarithms may be
obtained by expressing the summations (2.3-2.5) in the following form, in which
the dependence on $\lambda$ and $L$ factorizes in each term:
\begin{eqnarray}
\left( \begin{array}{cc}A(\lambda, L)\\B(\lambda, L)\\C(\lambda,L) \end{array}\right) = \sum_{n=0}^\infty 
\left( \begin{array}{cc}\rho_n(\lambda^2)\\\sigma_n(\lambda^2)\\\tau_n(\lambda^2) \end{array}\right) L^n,\nonumber\\
\left( \begin{array}{cc}\rho_n(\lambda^2)\\\sigma_n(\lambda^2)\\\tau_n(\lambda^2) \end{array}\right) = \sum_{m=n}^\infty 
\left( \begin{array}{cc}a_{m,n}\\b_{m,n}\\c_{m,n} \end{array}\right) (\lambda^2)^m.
\end{eqnarray}
Such a restructuring is developed into a procedure for closed form
summation in the Appendix to ref.\ [10]. We utilize this procedure in the
present section to demonstrate how $A(\lambda , L)$ within eq.\ (3.1) may be
expressed in closed form. If we substitute the expression (3.1) for
$A(\lambda , L)$ into the $RGE$ (2.10), we obtain the recursion relation
\begin{equation}
\left( 2 + \gamma_m (\lambda)\right) (n+1) \rho_{n+1} = \left( \beta(\lambda) \frac{\partial}{\partial \lambda} - 2\gamma_{\Gamma} (\lambda) \right) \rho_n .
\end{equation}
We now define
\begin{equation}
\rho_n (\lambda) \equiv \exp \left[ \int_0^\lambda \frac{2\gamma_\Gamma (w)}{\beta(w)}dw\right] \tilde{\rho}_n (\lambda)
\end{equation}
and find from eq.\ (3.2) that
\begin{equation}
\tilde{\rho}_{n+1} = \left( \frac{1}{n+1} \right) \left( \frac{1}{2+\gamma_m (\lambda)} \right) \beta(\lambda) \; \;  \frac{d \tilde{\rho}_n}{d\lambda} .
\end{equation}

From eqs.\ (3.3) and (3.4) it is straightforward to show that
\begin{equation}
\tilde{A} (\lambda, L) \equiv \sum_{n=0}^\infty \tilde{\rho}_n (\lambda) L^n = \exp \left[ \frac{L \beta(\lambda)}{2+\gamma_m (\lambda)} \frac{d}{d\lambda} \right] \tilde{\rho}_0 (\lambda).
\end{equation}
It is convenient here to introduce a new variable $\eta$ such that
\begin{equation}
\frac{\beta (\lambda)}{2+\gamma_m (\lambda)} \frac{d}{d\lambda} \equiv \frac{d}{d \eta}.
\end{equation}
This relation between differential operators is upheld provided
\begin{equation}
\eta (\lambda) = \int_{\lambda_0}^\lambda \frac{(2+\gamma_m (w))}{\beta(w)} dw.
\end{equation}
Equation (3.7) may be understood as defining $\lambda$ as an implicit
function of $\eta$ such that $f(\eta(\lambda)) = \lambda$. Upon
substitution of the differential operator (3.6) into eq.\ (3.5), we see that
\begin{eqnarray}
\tilde{A} (\lambda, L) & = & \exp \left( L \frac{d}{d\eta} \right) \tilde{\rho}_0 \left( f(\eta(\lambda))\right)\nonumber\\
& = & \tilde{\rho}_0 \left( f(\eta(\lambda) + L) \right).
\end{eqnarray}
We then obtain from eqs.\ (3.2-3.5) and (3.8) the following closed form solution for $A(\lambda, L)$:
\begin{eqnarray}
A(\lambda, L) & = & \exp \left[ \int_0^\lambda \frac{2\gamma_\Gamma (w)}{\beta(w)} dw \right] \tilde{A} (\lambda, L)\nonumber\\
& = & \exp \left[ \int_0^\lambda \frac{2\gamma_{\Gamma} (w)}{\beta(w)} dw \right] \tilde{\rho}_0 \left( f(\eta(\lambda)+L) \right)\nonumber\\
& = & \exp \left[ -\int_\lambda^{f(\eta(\lambda)+L)} \frac{2\gamma_{\Gamma} (w)}{\beta(w)} dw \right] \rho_0 \left( f(\eta(\lambda)+L) \right).
\end{eqnarray}
The employment of such an all-orders solution, of course, rests upon
complete knowledge of the function $\rho_0(\lambda)$, hence complete
knowledge of the coefficients $a_{m,0}$ for all $m$ (eq.\ (3.1)), as well
as knowledge of the inverse function $f$ of $\eta(\lambda)$, as defined by eq.\
(3.7). For this inverse function, it is useful to note that knowledge of
coefficients $a_{n,0}$, $b_{n,0}$, $c_{n,0}$ is sufficient to determine
the $n^{th}$-order coefficients of the $\beta$ and $\gamma$-functions (2.7-2.9)
appearing in eqs.\ (3.7) and (3.9); this is evident for the $n = 2$ case in
the derivation of the coefficients (2.24) and (2.25) for these functions. 

To illustrate how the solution (3.9) can actually be of use, we work with just 
leading order expressions $\beta(\lambda) =
B_3\lambda^3$, $\gamma_m(\lambda) = G_2\lambda^2$, and
$\gamma_\Gamma(\lambda) = D_2\lambda^2$, $\rho_0 = a_{0,0} = 1$,
and we find from eq.\ (3.7) that
\begin{equation}
-2B_3 \eta (\lambda) = \frac{2}{\lambda^2} - G_2 \log \frac{\lambda^2}{\lambda_0^2} - \frac{2}{\lambda_0^2}.
\end{equation}
To solve for $\lambda = f(\eta (\lambda))$, we define $W \equiv 2 / (G_2 \lambda^2)$ and find
from eq.\ (3.10) that
\begin{equation}
W + \log W = -\frac{2 B_3 \eta}{G_2} + \frac{2}{G_2 \lambda_0^2} + \log \left( \frac{2}{G_2 \lambda_0^2}\right)
\end{equation}
or
\begin{equation}
W e^W = \left[ \frac{2}{G_2 \lambda_0^2} \exp \left[ \frac{2}{G_2 \lambda_0^2} - \frac{2B_3 \eta}{G_2} \right] \right].
\end{equation}
The relation $W (\xi) e^{W(\xi)} = \xi$ defines the Lambert W-function implicitly \cite{17}, in which case
\begin{equation}
\lambda = f(\eta) = \left( \frac{2}{G_2 \mbox{\Large \it W} \left[ \frac{2}{G_2 \lambda_0^2} \exp \left( \frac{2}{G_2 \lambda_0^2} - \frac{2B_3 \eta}{G_2} \right) \right]} \right)^{1/2}.
\end{equation}
We substitute $f(\eta + L)$, as defined by eq.\ (3.13), directly into eq.\ (3.9) to obtain
\begin{eqnarray}
A(\lambda, L) & = & \exp\left[ -\int_{\lambda}^{f(\eta + L)} \frac{2 D_2 w^2}{B_3 w^3} dw \right] \cdot 1\nonumber\\
& = & \left[ \frac{G_2 \lambda^2 \mbox{\Large \it W} \left[ \frac{2}{G_2 \lambda_0^2} \exp \left( \frac{2}{G_2 \lambda_0^2} -\frac{2B_3 (\eta + L)}{G_2} \right) \right]}{2} \right]^{D_2/B_3}.
\end{eqnarray}
We then substitute eq.\ (3.10) into the final line of eq.\ (3.14) to obtain
\begin{equation}
A(\lambda, L) = \left[ \frac{G_2 \lambda^2}{2} \mbox{\Large \it W} \left[ \frac{2}{G_2 \lambda^2} \exp \left( \frac{2}{G_2 \lambda^2} - \frac{2 B_3 L}{G_2} \right)\right] \right]^{D_2/B_3}.
\end{equation}

Eq.\\ (3.15) is a closed form solution to the RGE (2.10) based upon the premise that $B(\lambda)$, $\gamma_m (\lambda)$ and $\gamma_\Gamma (\lambda)$ are given 
entirely by their leading order terms:  $B_3 \lambda^3$, $G_2 \lambda^2$, and $D_2 \lambda^2$, respectively.  In Appendix
A, we compare eq.\ (3.15) with the expression
\begin{equation}
A(\lambda, L) \rightarrow R_0 (\lambda^2 L) = (1 - B_3 \lambda^2 L)^{D_2/B_3}
\end{equation}
derived from equivalent lowest-order assumptions in the previous section.  Specifically, we show that eqs.\ (3.15)
and (3.16) differ at most by ${\cal{O}}(\lambda^4)$, as expected.\footnote{If they differed by 
${\cal{O}}(\lambda^2)$, then $B_3$ in eq.\ (3.16) could be replaced by a new constant $B_3^{\prime}$ to eliminate
the ${\cal{O}}(\lambda^2)$ discrepancy, in which case eq.\ (3.15) would not be a ``best fit'' to the expression (3.16).}

\newpage


\section{RG-Resummation of the Yang-Mills Effective Action}

\renewcommand{\theequation}{4.\arabic{equation}}
\setcounter{equation}{0}

The analysis of Sections 2 and 3 is also applicable to Yang-Mills theory.  
If we employ background field quantization 
\cite{18}, then the renormalized effective action for Yang-Mills
theory involves the following gauge-invariant form
\begin{equation}
\Gamma[A] = -\frac{1}{4} F_{\mu\nu}^a F^{\mu\nu a} S [L, g^2]
\end{equation}
where in configuration space
\begin{equation}
F_{\mu\nu}^a = \partial_\mu A_\nu^a - \partial_\nu A_\mu^a + gf^{abc}
A_\mu^b A_\nu^c
\end{equation}
and where
\begin{equation}
L \equiv \log (\mu^2 / p^2)
\end{equation}
with $p$ being an external momentum, and $\mu$ being a renormalization
scale parameter.  We find analogous to eq.\ (2.6) that the effective
action is invariant under changes in the unphysical parameter $\mu$:
\begin{equation}
0 = \mu^2 \frac{d\Gamma}{d\mu^2} = \left( \mu^2 \frac{\partial}{\partial \mu^2} + \beta (g^2)
\frac{\partial}{\partial g^2} + \frac{\gamma(g^2)}{2} \int d^4 x
A_\eta^a (x) \frac{\delta}{\delta A_\eta^a (x)} \right) \Gamma [A]
\end{equation}
\begin{equation}
\beta(g^2) = \sum_{k=2}^\infty g^{2k} \; b_k \equiv \mu^2 \frac{d g^2
(\mu)}{d\mu^2}
\end{equation}
\begin{equation}
\frac{\gamma(g^2)}{2} A_\eta^a = \mu^2 \frac{\partial A_\eta^a}{\partial
\mu^2} (x, \mu).
\end{equation}

Two features of Yang-Mills theory make the effective-action (4.1) more
tractable than that of the $\phi_6^3$ model considered in Section 2.
First of all, $g A_\eta^a$ is unrenormalized in order to
preserve gauge invariance in the background field if gauge fixing is
chosen appropriately \cite{18,19}, in which case
\begin{equation}
0 = \mu^2 \frac{d}{d\mu^2} \left[ g(\mu) A_\eta^a (x, \mu) \right] =
\left( \frac{\beta(g^2)}{2g} \right) A_\eta^a (x, \mu) + g \left(
\frac{\gamma(g^2)}{2} A_\eta^a (x, \mu) \right).
\end{equation}
Consequently we find that
\begin{equation}
\gamma(g^2) = -\beta (g^2) / g^2.
\end{equation}
Secondly, we note that the logarithm $L$ in the effective action (4.1)
has only explicit dependence on the renormalization mass scale $\mu^2$;
by contrast the logarithm appearing in the effective action (2.2) has
both explicit and implicit (through $m(\mu)$) dependence on $\mu^2$.

Upon substitution of eqs.\ (4.1) and (4.8) into the RG-equation (4.4), we 
obtain
\begin{equation}
\left[\frac{\partial}{\partial L} + \beta (g^2) \left(
\frac{\partial}{\partial g^2} - \frac{1}{g^2} \right) \right] S [L, g^2]
= 0.
\end{equation}
As before, we can expand the effective action's scalar function as a
double summation in $L$ and $g^2$
\begin{equation}
S [L, g^2] = \sum_{n=0}^\infty \sum_{m=0}^n y_{n,m} g^{2n} L^m.
\end{equation}
Upon substitution of this series and the $\beta$-function series (4.5)
into the RG-equation (4.9), we find that the aggregate coefficient of
$g^{2p} L^{p-k}$ necessarily vanishes (for positive integer values of
$p$ and $k$ with $\; k < p$):
\begin{equation}
(p - k + 1)y_{p, p-k+1} + \sum_{\ell=1}^k (p - \ell - 1) b_{\ell+1} \;  y_{p -
\ell, p-k} = 0.
\end{equation}
The double summation (4.10) can, as before, be rearranged into the form
\begin{equation}
S [L, g^2] = \sum_{n=0}^\infty g^{2n} U_n (g^2 L)
\end{equation}
where $(v = g^2 L)$
\begin{equation}
U_n (v) \equiv \sum_{k=n}^\infty y_{k, k-n} v^{k-n}.
\end{equation}
The recursion relations (4.11) can be used to obtain an explicit
set of
 first order linear differential equations for $U_n (v)$.  If one multiplies 
(4.11) by $v^{p-k}$ and then sums over $p$ from $p = k$ to infinity $(k \geq 1)$,
one finds that
\begin{eqnarray}
0 & = & \sum_{p=k}^\infty (p - (k-1)) y_{p, p-(k-
1)} v^{p-k}\nonumber\\
& + & \sum_{\ell=1}^k b_{\ell + 1} \left( v \sum_{p=k}^\infty (p-k)
y_{p-\ell, p-k} v^{p-k-1} + (k - \ell - 1) \sum_{p=k}^\infty y_{p-\ell, p-k} v^{p-k}
\right)\nonumber\\
& = & \frac{d U_{k-1}}{d v} + \sum_{\ell=1}^k b_{\ell + 1} \left(
v \frac{d U_{k-\ell}}{d v} + (k - \ell - 1) U_{k - \ell} \right).
\end{eqnarray}
Given the initial conditions
\begin{equation}
U_n (0) = y_{n,0}
\end{equation}
one may solve the differential equations (4.14) successively for $U_0$,
$U_1$, $U_2$ etc.  For example, the solutions to these equations
accessible from the set of known coefficients $\{b_2, b_3, b_4, b_5 \}$ of the 
Yang-Mills $\beta$-function are
\begin{equation}
U_0 (v) = y_{0,0} (1 + b_2 v)
\end{equation}
\begin{equation}
U_1 (v) = y_{1,0} + \frac{b_3}{b_2} y_{0,0} \log (1 + b_2 v)
\end{equation}
\begin{equation}
U_2 (v) = \frac{y_{2,0} + y_{0,0} \left[ \frac{b_3^2}{b_2^2} \log (1 +
b_2 v) + \left( b_4 - \frac{b_3^2}{b_2} \right) v\right]}{(1 + b_2 v)}
\end{equation}
\begin{eqnarray}
U_3 (v) & = & \left[ y_{3,0} - y_{2,0} \frac{b_3}{b_2} \log (1 + b_2 v)
\right] / (1 + b_2 v)^2\nonumber\\
& + & y_{0,0} \Biggl\{ \left[ \frac{b_5}{2b_2} - \frac{b_3 b_4}{b_2^2} +
\frac{b_3^3}{2b_2^3} \right] + \left[ \frac{b_3 b_4}{b_2^2} -
\frac{b_3^3}{b_2^3} \right] / (1 + b_2 v) \Biggr.\nonumber\\
& + & \Biggl. \frac{ \left[ \frac{b_3^3}{2b_2^3} - \frac{b_5}{2b_2} + \frac {b_3
b_4}{b_2^2} \log (1 + b_2 v) - \frac{b_3^3}{2b_2^3} \log^2 (1 + b_2 v)
\right]}{(1 + b_2 v)^2} \Biggr\}
\end{eqnarray}
where the $\overline{MS}$ $SU(N)$ Yang-Mills $\beta$-function
coefficients, as defined by eq.\ (4.5) are \cite{20}
\begin{eqnarray}
b_2 = -\frac{11 N}{12(4\pi^2)}, \; \; b_3 = -\frac{17
N^2}{24(4\pi^2)^2}, \; \; b_4 = -\frac{2857
N^3}{3456(4\pi^2)^3},\nonumber\\
b_5 = -\frac{\left( N^4 \left[ \frac{150473}{486} + \frac{44}{9} \zeta
(3) \right] + N^2 \left[ -\frac{40}{3} + 352 \zeta (3) \right]
\right)}{256 (4\pi^2)^4}.
\end{eqnarray}

The utility of the solution (4.12) is limited, however, by our 
ignorance of the coefficients $y_{k,0}$ appearing in eqs.\ (4.16)-(4.19).
It is possible to achieve deeper insight into the behaviour of $S[L, g^2]$ using
the methods of Section 3.  As before, we expand the
effective Lagrangian (4.1) in powers of $L$:
\begin{equation}
S[L, g^2] \equiv \sum_{n=0}^\infty B_n (g^2) L^n
\end{equation}
If we substitute this series into the RG equation (4.9) we find that 
\begin{equation}
(n+1) B_{n+1} + g^2 \beta(g^2) \frac{d}{dg^2} \left( \frac{B_n
(g^2)}{g^2} \right) = 0
\end{equation}
Let $\tilde{B}_n (g^2) \equiv B_n (g^2)/g^2$, in which case
\begin{equation}
\tilde{B}_{n+1} (g^2) = -\frac{\beta(g^2)}{(n+1)} \frac{d}{dg^2}
\tilde{B}_n (g^2) .
\end{equation}
If we define a variable $y$ implicitly via the relation
\begin{equation}
\beta(g^2) \frac{d}{d g^2} \equiv \frac{d}{dy}
\end{equation}
we see from eq.\ (4.23) that
\begin{equation}
\tilde{B}_{n+1} (y) = -\frac{1}{(n+1)} \frac{d}{dy} \tilde{B}_n (y),
\end{equation}
in which case
\begin{eqnarray}
S[L, g^2] & = & g^2 (\mu) \sum_{n=0}^\infty \tilde{B}_n L^n\nonumber\\
& = & g^2(\mu) \sum_{n=0}^\infty \left( -\frac{1}{n!} L^n
\frac{d^n}{dy^n} \right) \tilde{B}_0 \left( g^2 [y] \right)\nonumber\\
& = & g^2 (\mu) \exp \left[ -L \frac{d}{dy} \right] \tilde{B}_0 \left(
g^2 [y] \right)\nonumber\\
& = & g^2 (\mu) \tilde{B}_0 \left( g^2 [y - L] \right) .
\end{eqnarray}
Now the definition (4.24) implies that
\begin{equation}
y \left[g^2 (\mu) \right] = \int_{g^2 (p)}^{g^2 (\mu)}
\frac{d\xi}{\beta(\xi)} .
\end{equation}
Note from eq.\ (4.27) that the implicit function $y$ has been chosen
such that $g^2 [y] = g^2(\mu)$, $g^2[0]=g^2(p)$; i.e. $y[g^2(p)]=0$.
However, the defining relation (4.5) for the $\beta$-function necessarily implies that
\begin{equation}
L \equiv \log \left( \frac{\mu^2}{p^2} \right) = \int_{g^2(p)}^{g^2(\mu)} \frac{ds}{\beta(s)}.
\end{equation}
Consequently, we see from eqs.\ (4.27) and (4.28) that $y = L$, in which case
we find from eq.\ (4.26) that
\begin{eqnarray}
S[L, g^2] & = & g^2 (\mu) \; \tilde{B}_0 \left[ g^2 [0]\right]\nonumber\\
& = & g^2 (\mu) \; \tilde{B}_0 \left[ g^2 (p) \right].
\end{eqnarray}

Since $g^2 \tilde{B}_0 (g^2) = B_0 (g^2) = \sum_{n=0}^\infty y_{n,0}
g^{2n}$, we find that
\begin{equation}
S[L, g^2] = \left( \sum_{n=0}^\infty y_{n,0} \; \; \left[ g(p)\right]^{2(n-1)} \right) g^2
(\mu).
\end{equation}
We thus see from eqs.\ (4.1) and (4.30) that the effective Lagrangian is proportional
to $g^2 (\mu) F_{\mu\nu}^a F^{\mu\nu a}$.  Note from eq.\ (4.30) that $S = k g^2 (\mu)$, where
$k$ is independent of $\mu$, and that such a result is a valid solution to the 
RG-equation (4.9).\footnote{The result $S = k g^2 (\mu)$ can be extracted trivially 
from eqs.\ (4.1) and (4.2) by noting that the scale invariance of 
$g(\mu) A_\nu^a (x, \mu)$ [eq.\ (4.7)], implies concomitant scale invariance of 
$g(\mu) F_{\mu\nu}^a$, and consequently, of $g^2 (\mu) F_{\mu\nu}^a F^{\mu\nu a}$.}  

The solution (4.30) is particularly useful in theories in which the $\beta$-function
is known to all orders. Both the form (4.1) for the logarithm-dependent portion of the bosonic 
contribution to the effective Lagrangian and the RG-equation (4.9) remain applicable to $N = 1$ 
supersymmetric Yang-Mills theory. For this supersymmetric case, however, the 
$\beta$-function can be obtained to all orders by requiring that the supermultiplet
structure of the theory uphold the Adler-Bardeen theorem \cite{15}.  Indeed, the 
$\beta$-function obtained by this method is found
to agree with that extracted by instanton calculus methods \cite{14};  for the
$SU(3)$ case this $\beta$-function is found to be
\begin{equation}
\beta(g^2) = \mu^2 \frac{dg^2}{d\mu^2} = -\frac{9g^4/16\pi^2}{1-3g^2/8\pi^2} ,
\end{equation}
in which case the couplant $x(\mu) \equiv g^2 (\mu) / 16\pi^2$ satisfies
the constraint
\begin{equation}
-\frac{1}{6x(\mu)} \exp \left[ -\frac{1}{6x(\mu)} \right] = -
\frac{1}{6x(p)} \left( \frac{\mu^2}{p^2} \right)^{1/6} \exp \left[ -
\frac{1}{6x(p)} \right].
\end{equation}
Given the defining relationship $f e^f = \xi$ for the Lambert W-function 
$f = W[\xi]$, we obtain from eq.\ (4.32) the following closed-form 
expression for the $F^2$-dependent portion of the effective Lagrangian (4.30) of supersymmetric $N_c = 3$ 
Yang-Mills theory:
\begin{equation}
\Gamma [A] = \frac{2\pi^2 \left( \sum_{n=0}^\infty y_{n,0} \; \;  \left[g(p)\right]^{2(n-1)}
\right) F_{\mu\nu}^a (A) F^{\mu\nu a} (A)}{3 \mbox{\Large \it W} \left[ -
\frac{8\pi^2}{3g^2 (p)} \left( \frac{\mu^2}{p^2} \right)^{1/6} \exp
\left( -\frac{-8\pi^2}{3g^2(p)} \right) \right]}
\end{equation}

\newpage


\section{The RG-Invariant Effective Couplant}

\renewcommand{\theequation}{5.\arabic{equation}}
\setcounter{equation}{0}

An unanticipated consequence of the previous section is a new and useful
series representation for the RG-invariant couplant $x(p) \equiv g^2 (p)
/ 4\pi^2$ (for QCD $x(\sqrt{s}) = \alpha_s (\sqrt{s})/4\pi$) that does not
involve an order-by-order redefinition of the scale
parameter $\Lambda$.  To obtain this series, we first note that 
eqs.\ (4.16)-(4.19) and eq.\ (4.30) are derived for arbitrary values of
$y_{n,0}$.  We further note that if
one were to choose $y_{n,0} = \delta_{n,0}$, then the expression (4.30)
would just be $g^2 (\mu) / g^2 (p)$.  However, if we make the choice $y_{n,0} =
\delta_{n,0}$ in eqs.\ (4.16)-(4.19), the series expansion (4.12)
corresponding to the same quantity as eq.\ (4.30) is seen to be 
\begin{eqnarray}
\frac{g^2(\mu)}{g^2(p)} & = & w + g^2 (\mu) \frac{b_3}{b_2} \log
(w)\nonumber\\
& + & \frac{g^4(\mu)}{w} \left[ \frac{b_3^2}{b_2^2} \log(w) + \left(
\frac{b_4}{b_2} - \frac{b_3^2}{b_2^2} \right) (w - 1) \right]\nonumber\\
& + & \frac{g^6(\mu)}{w^2} \left[ \left( \frac{b_5}{2b_2} - \frac{b_3
b_4}{b_2^2} + \frac{b_3^3}{2b_2^3} \right) w^2 \right.\nonumber\\
& + & \left( \frac{b_3 b_4}{b_2^2} - \frac{b_3^3}{b_2^3} \right) w +
\left( \frac{b_3^3}{2b_2^3} - \frac{b_5}{2b_2} \right)\nonumber\\
& + & \left. \frac{b_3 b_4}{b_2^2} \log (w) - \frac{b_3^3}{2b_2^3} \log^2 (w) \right]
+ {\cal{O}} \left[ g^8 (\mu) \right] ,
\end{eqnarray}
where
\begin{equation}
w \equiv 1 + b_2 g^2 (\mu) \log \left( \frac{\mu^2}{p^2} \right) ,
\end{equation}
and where the constants $b_k$ are as defined by the $\beta$-function
(4.5).  For our purposes here, we will utilize the (somewhat more)
standard conventions of QCD:
\begin{equation}
x(\mu) = \alpha_s(\mu)/\pi = g^2 (\mu) / 4\pi^2
\end{equation}
\begin{equation}
\mu^2 \frac{d}{d\mu^2} x(\mu) = -\sum_{k=0}^\infty \beta_k \left( x(\mu)
\right)^{k+2}.
\end{equation}
Comparing eqs.\ (4.5) and (5.4), we see that
\begin{equation}
b_{k+2} = - \beta_k / (4\pi^2)^{k+1}
\end{equation}
and that
\begin{equation}
w = 1 - \beta_0 x(\mu) \log (\mu^2/p^2).
\end{equation}
Using the definition (5.3) and the $\beta$-function coefficients defined
by eqs.\ (5.4) and (5.5), we find from eq.\ (5.1) that
\begin{eqnarray}
\frac{x(\mu)}{x(p)} & = & w + x(\mu) \frac{\beta_1}{\beta_0} \log
(w)\nonumber\\
& + & \frac{x^2(\mu)}{w} \left[ \frac{\beta_1^2}{\beta_0^2} \log (w) +
\left( \frac{\beta_2}{\beta_0} - \frac{\beta_1^2}{\beta_0^2} \right) (w
- 1) \right]\nonumber\\
& + & \frac{x^3(\mu)}{w^2} \left[ \left( \frac{\beta_3}{2\beta_0} -
\frac{\beta_1 \beta_2}{\beta_0^2} + \frac{\beta_1^3}{2\beta_0^3} \right)
w^2 \right.\nonumber\\
& + & \left( \frac{\beta_1 \beta_2}{\beta_0^2} - \frac{
\beta_1^3}{\beta_0^3} \right) w + \left( \frac{\beta_1^3}{2\beta_0^3} -
\frac{\beta_3}{2\beta_0} \right)\nonumber\\
& + & \left. \frac{\beta_1 \beta_2}{\beta_0^2} \log (w) -
\frac{\beta_1^3}{2\beta_0^3} \log^2 (w) \right]\nonumber\\
& + & {\cal{O}}\left( x^4 (\mu) \right),
\end{eqnarray}
with $w$ given by eq.\ (5.6).

Now the expression we seek to obtain is an expression for the 
RG-invariant couplant $x(p)$, where $p$ is an external physical momentum
scale, in terms of a chosen reference momentum scale $\mu$ and a
reference couplant value $x(\mu)$.  As an obvious example, the solution
of eq.\ (5.4) when only the leading one-loop (1L) coefficient $\beta_0$
contributes to the right-hand side is just
\begin{equation}
x_{1L} (p) = \frac{x(\mu)}{1 - \beta_0 x(\mu) \log \left(
\frac{\mu^2}{p^2} \right)} = \frac{x(\mu)}{w}
\end{equation}
For the full $\beta$-function (5.4), our solution must necessarily be of
the form
\begin{equation}
x(p) = x(\mu) \left[ S_0 + x(\mu) S_1 + x^2 (\mu) S_2 + x^3 (\mu) S_3 +
... \right],
\end{equation}
where the coefficients $S_0$, $S_1$, $S_2$, $S_3$ are functions of
$x(\mu)$ and $w$.  To find these coefficients algebraically, we simply
require that the product  of $x(\mu) / x(p)$, as given by eq.\ (5.7), and
$x(p) / x(\mu)$ as given by eq.\ (5.9), be
equal to unity on an order-by-order basis in $x(\mu)$:
\begin{equation}
w S_0 = 1
\end{equation}
\begin{equation}
w S_1 + S_0 \frac{\beta_1}{\beta_0} \log (w) = 0
\end{equation}
\begin{equation}
w S_2 + S_1 \frac{\beta_1}{\beta_0} \log (w) + \frac{S_0}{w} \left[
\frac{\beta_1^2}{\beta_0^2} \log (w) + \left( \frac{\beta_2}{\beta_0} -
\frac{\beta_1^2}{\beta_0^2} \right) (w - 1) \right] = 0
\end{equation}
\begin{eqnarray}
w S_3 & + & S_2 \frac{\beta_1}{\beta_0} \log (w) + \frac{S_1}{w} \left[
\frac{\beta_1^2}{\beta_0^2} \log (w) + \left( \frac{\beta_2}{\beta_0} -
\frac{\beta_1^2}{\beta_0^2} \right) (w - 1) \right]\nonumber\\
& + & \frac{S_0}{w^2} \left[ \left( \frac{\beta_3}{2\beta_0} -
\frac{\beta_1 \beta_2}{\beta_0^2} + \frac{\beta_1^3}{2\beta_0^3} \right)
w^2 + \left( \frac{\beta_1 \beta_2}{\beta_0^2} -
\frac{\beta_1^3}{\beta_0^3} \right) w \right.\nonumber\\
& + & \left. \left( \frac{\beta_1^3}{2\beta_0^3} -
\frac{\beta_3}{2\beta_0} \right) + \frac{\beta_1 \beta_2}{\beta_0^2}
\log (w) - \frac{\beta_1^3}{2\beta_0^3} \log^2 (w) \right] = 0.
\end{eqnarray}
By solving these equations sequentially, and the substituting into eq.\
(5.9), we find that
\begin{eqnarray}
x(p) & = & \frac{x(\mu)}{w} - \frac{x^2 (\mu)}{w^2} \frac{\beta_1}{\beta_0} 
        \log (w)\nonumber\\
& + & \frac{x^3(\mu)}{w^3} \left[ \left( \frac{\beta_1^2}{\beta_0^2} 
        - \frac{\beta_2}{\beta_0} \right) (w - 1) - \frac{\beta_1^2}{\beta_0^2} 
        \left( \log (w) - \log^2 (w) \right) \right]\nonumber\\
& + & \frac{x^4 (\mu)}{w^4} \left[ \left( -\frac{\beta_3}{2\beta_0} +
        \frac{\beta_1 \beta_2}{\beta_0^2} - \frac{\beta_1^3}{2\beta_0^3} \right)
        w^2 + \left( \frac{\beta_1^3}{\beta_0^3} - \frac{\beta_1
        \beta_2}{\beta_0^2} \right) w \right.\nonumber\\
& + & \left( \frac{2\beta_1 \beta_2}{\beta_0^2} -
        \frac{2\beta_1^3}{\beta_0^3} \right) w \log (w)\nonumber\\
& + & \left( \frac{\beta_3}{2\beta_0} - \frac{\beta_1^3}{2\beta_0^3}
        \right) + \left( \frac{2\beta_1^3}{\beta_0^3} - \frac{3\beta_1
        \beta_2}{\beta_0^2} \right) \log (w)\nonumber\\
& + & \left. \frac{\beta_1^3}{\beta_0^3} \left( \frac{5}{2} \log^2 (w) - \log^3
        (w) \right) \right] + {\cal{O}} \left( x^5 (\mu) \right).
\end{eqnarray}
The leading three orders of the above equation have been derived previously (see eq.\ (16)
of ref.\ [12]). In Appendix B, we derive eq.\ (5.14) via an alternative procedure utilizing
the explicit RG-invariance of the couplant $x(p)$.
This expression may be interpreted as an expression for the four-loop
effective couplant in terms of $x(\mu)$ and the one-loop couplant
$x_{1L} (p)$, as given by eq.\ (5.8):
\begin{eqnarray}
x_{4L} (p) & = & x_{1L} (p) - \frac{\beta_1}{\beta_0} x_{1L}^2 (p) \log \left( \frac{x(\mu)}{x_{1L} (p)} \right)\nonumber\\
& + & x_{1L}^3 (p) \left[ \left( \frac{\beta_1^2}{\beta_0^2} - \frac{\beta_2}{\beta_0} \right) \left( \frac{x(\mu)}{x_{1L} (p)} - 1 \right)\right.\nonumber\\
& - & \left. \frac{\beta_1^2}{\beta_0^2} \left( \log \left( \frac{x(\mu)}{x_{1L}(p)} \right) - \log^2 \left( \frac{x(\mu)}{x_{1L}(p)} \right) \right] \right.\nonumber\\
& + & x_{1L}^4 (p) \left[ \left( -\frac{\beta_3}{2\beta_0} + \frac{\beta_1 \beta_2}{\beta_0^2} - \frac{\beta_1^3}{2\beta_0^3} \right) \left( \frac{x(\mu)}{x_{1L} (p)} \right)^2 \right.\nonumber\\
& + & \left( \frac{\beta_1^3}{\beta_0^3} - \frac{\beta_1 \beta_2}{\beta_0^2} \right) \left( \frac{x(\mu)}{x_{1L} (p)} \right)\nonumber\\
& + & \left( \frac{2\beta_1 \beta_2}{\beta_0^2} - \frac{2\beta_1^3}{\beta_0^3} \right) \left( \frac{x(\mu)}{x_{1L} (p)} \right) \log \left( \frac{x(\mu)}{x_{1L} (p)} \right)\nonumber\\
& + & \left( \frac{\beta_3}{2\beta_0} - \frac{\beta_1^3}{2\beta_0^3} \right) + \left( \frac{2\beta_1^3}{\beta_0^3} - \frac{3\beta_1 \beta_2}{\beta_0^2} \right) \log \left( \frac{x(\mu)}{x_{1L} (p)} \right)\nonumber\\
& + & \left. \frac{\beta_1^3}{\beta_0^3} \left( \frac{5}{2} \log^2 \left( \frac{x (\mu)}{x_{1L} (p)} \right) - \log^3 \left( \frac{x(\mu)}{x_{1L} (p)} \right) \right) \right]\nonumber\\
& + & {\cal{O}} \left( x_{1L}^5 (p) \right).
\end{eqnarray}

In Fig. 1, we display for $n_f = 3$ QCD a comparison of the running
couplant (5.14) to the corresponding explicit (numerical) solution to
the differential equation (5.4) truncated after the last known
contribution $(-\beta_3 x^5)$.  To obtain these plots, we assume that
$\mu = m_\tau$ and that $x(m_\tau) = \alpha_s (m_\tau) / \pi =
0.33/\pi$ \cite{21}. The two curves are seen to coincide until $\sqrt{p^2} < 1$ GeV, 
in which case the expression (5.14) falls below the true couplant value.
This discrepancy follows from the fact that the series (5.14) is 
term-by-term singular at $w = 0$, corresponding to the Landau pole of the
one-loop couplant (5.8).  However, the Landau pole of the true 4-loop
couplant, as evolved directly from the differential equation (5.4), is
necessarily above the $w = 0$ one-loop pole at $p^2 = m_{\tau}^2 \exp
[-1/\beta_0 x(m_{\tau})]$ because $\{\beta_0, \beta_1, \beta_2, \beta_3
\}$ are all positive;  {\it i.e.} the four-loop $\beta$-function drives the couplant
more quickly to large values than the one-loop $\beta$-function.

This behaviour becomes somewhat more transparent if we specialize to
$n_f = 3$ QCD in the 't Hooft renormalization scheme \cite{22} in which
$\beta_0$ and $\beta_1$ have their $\overline{MS}$ values, but all
subsequent $\beta$-function coefficients are zero.  In Fig. 2 we display
successively higher order plots of the series (5.9) in the 't Hooft
scheme with the same initial conditions at $m_\tau$ as in Fig. 1.  The
figure displays the lowest-order (1L) approximation 
$x(p) \cong x(\mu) S_0$, the next-to-lowest (NL) order approximation
$x(p) \cong x(\mu)
\left( S_0 + x(\mu) S_1 \right)$, the subsequent (NNL) approximation
$x(p) \cong x(\mu) \left( S_0 + x(\mu) S_1 + x^2 (\mu) S_2 \right)$, and
finally the NNNL approximation $x(p) \cong x(\mu) \left( S_0 + x(\mu)
S_1 + x^2 (\mu) S_2 + x^3 (\mu) S_3 \right)$.  The functions $S_0$ and
$S_1$ are as given in eq.\ (5.14) [or alternatively, eqs.\ (B.16) and
(B.19) of Appendix B].  The functions $S_2$ and $S_3$ are also given in
eq.\ (5.14) [or eqs.\ (B.22) and (B.23)], but with $\beta_2$ and $\beta_3$
in these equations taken to be zero, consistent with the 't Hooft
renormalization scheme.  Since the lowest order approximation is just the
one-loop running couplant (5.8), the figure shows how the incorporation
of successive terms in the series (5.9) takes one from the 
one-loop couplant $x_{1L} (p)$ to approximations that grow
successively closer to the true 't Hooft scheme ('tH) couplant,
defined as the solution of eq.\ (5.4) with $\beta_k = 0$ for $k \geq 2$.

However, it is also evident from the figure that {\it all} the series
approximations listed above differ from the true solution in the infrared 
region.  Each approximation to the series (5.9) is term-by-term
singular at $w = 0$, the Landau pole of $x_{1L} (p)$ [eq.\ (5.8)] which
occurs (for $\mu = m_\tau$) at
\begin{equation}
p^2 = m_{\tau}^2 \exp \left[ -1 / \beta_0 x (m_\tau) \right].
\end{equation}
By contrast, the exact 't Hooft scheme solution to eq.\ (5.4) satisfies
the constraint
\begin{eqnarray}
\beta_0 \log \left( \frac{m_{\tau}^2}{p^2} \right) & = &
        \frac{1}{x(m_\tau)} - \frac{1}{x(p)}\nonumber\\
& + & \frac{\beta_1}{\beta_0} \log \left[ \frac{x(m_\tau) [x(p) +
        \beta_0 / \beta_1]}{x(p)[x(m_\tau) + \beta_0 / \beta_1]} \right] .
\end{eqnarray} 
The Landau pole of eq.\ (5.17) occurs when $x(p) \rightarrow \infty$, {\it i.e.}, when
\begin{equation}
p^2 = m_{\tau}^2 \exp \left[ -1 / \beta_0 x(m_\tau) \right] \left[ 1 +
\frac{\beta_0}{\beta_1 x(m_\tau)} \right]^{\frac{\beta_1}{\beta_0^2}}.
\end{equation}
Since $\beta_0$ and $\beta_1$ are both positive, the true 't 
Hooft-scheme Landau pole (5.18) is clearly larger than the singularity (5.16)
characterizing each term of the series (5.9).  Thus, the true 't Hooft
scheme couplant necessarily evolves faster than {\it any} approximation
based upon the series (5.9).

Indeed, the true couplant has an infrared bound (5.18) on its domain
that is well above the corresponding bound (5.16) on the terms of the
series (5.9).  Such a discrepancy between a series representation of a
function and the function itself is not unprecedented.  An illustrative
toy example is the function 
\begin{equation}
f(s) = (s - \Lambda^2) / (s - \Lambda^2 + x(s)),
\end{equation}
which is singular only at values of $s$ satisfying the constraint $s =
\Lambda^2 - x(s)$.  A perturbative expansion of this same function in
the expansion parameter $x(s)$,
\begin{equation}
f(s) = 1 - \frac{x(s)}{s-\Lambda^2} + \frac{x^2 (s)}{(s - \Lambda^2)^2} -
\frac{x^3 (s)}{(s - \Lambda^2)^3} \pm ...
\end{equation}
is term-by-term singular at $s = \Lambda^2$, as well as at any ``Landau
pole'' values of $s$ at which $x(s)$ is itself singular.  The analogy to
calculable perturbative processes is inescapable;  the occurrence of
term-by-term singularities at some value of $s$ in a perturbative field-
theoretical series for a physical process does not necessarily imply
that the process itself is inaccessible at that value of $s$.
Restrictions, such as a Landau pole, on the kinematical domain of the
running couplant, do not necessarily correspond to infrared
restrictions on the kinematical domain of the process itself.


\section{Instanton Contribution to the $SU(2)$ Effective Lagrangian}

\renewcommand{\theequation}{6.\arabic{equation}}
\setcounter{equation}{0}

Consider the instanton contribution to the effective Lagrangian of an $SU(2)$ gauge theory, 
as derived by 't Hooft \cite{16}.  This contribution contains an
integral over the instanton size $\rho$
\begin{equation}
{\cal{L}}_{eff} \sim  K \int d\rho \; \rho^{-5+3n_f} \exp \left\{ -
\frac{8\pi^2}{g^2(\mu)}S\right\}
\end{equation}
where the series $S$ is a power series in the scale dependent coupling
constant of the form
\begin{equation}
S = 1 + \sum_{n=1}^\infty \sum_{m=0}^n T_{n,m} g^{2n} (\mu) \log^m
(\mu \rho).
\end{equation}
The contribution to the measure $K$ in eq.\ (6.1) is independent of the
renormalization scale parameter $\mu$.
In ref.\ [16], the one loop contributions $T_{1,0}$ and $T_{1,1}$ to eq.\ (6.2)
are utilized explicitly to show that the integral (6.1) converges in the
ultraviolet limit $(\rho \rightarrow 0)$ but diverges in the infrared
limit $(\rho \rightarrow \infty)$.  In this section, we explore whether
the summation of successively subleading logarithms affects such
asymptotic behaviour.

We begin by regrouping the series (6.2) in terms of sequential
summations of leading $(S_0)$, next-to-leading $(S_1)$, 
next-to-next-to-leading $(S_2)$, ... , logarithmic terms
\begin{eqnarray}
S & = & \sum_{k=0}^\infty g^{2k} (\mu) \left( \sum_{\ell = k}^\infty
T_{\ell, \ell-k} \left[ g^2 (\mu) \log (\mu\rho)\right]^{\ell-k}
\right)\nonumber\\
& \equiv & \sum_{k=0}^\infty g^{2k} (\mu) S_k \left[ g^2 (\mu) \log
(\mu\rho)\right].
\end{eqnarray}
The RG-equation for the argument of the exponent occurring in the integrand of 
eq.\ (6.1) is just
\begin{equation}
0 = \left( \mu \frac{\partial}{\partial \mu} + \beta (g)
\frac{\partial}{\partial g} \right) \sum_{k = 0}^\infty \left[
g(\mu)\right]^{2(k-1)} S_k \left[ g^2 (\mu) \log (\mu \rho)\right],
\end{equation}
where $\beta(g)$ is the $SU(2)$ $\beta$-function with fermionic and
scalar field (s.f.) contributions:
\begin{equation}
\mu \frac{dg}{d\mu} = \beta(g) = \sum_{k=1}^\infty b_{2k+1} g^{2k+1}
\end{equation}
\begin{equation}
b_3 = -\left( \frac{22}{3} - \frac{2n_f}{3} - \mbox{s.f.}\right) /
16\pi^2.
\end{equation}
If one employs the series definition (6.3) for the $S_k$'s occurring on
the right hand side of (6.4), one can obtain recursion relations for the
$T_{n,m}$ coefficients.  One finds from $S_0$, for example, that the
aggregate coefficient of $g^{2(n-1)} L^{n-1}$ on the right hand side of
(6.4) will vanish only if
\begin{equation}
n T_{n,n} + b_3 (2n-4) T_{n-1, n-1} = 0.
\end{equation}
This recursion relation not only implies that $T_{1,1} = 2b_3$, a result
explicitly obtained in ref.\ [16],\footnote{The RG-consistency of this
result is also noted in ref.\ [16].} but also implies that $T_{n,n}=0$ for $n \geq 2$.  
Consequently, the summation of leading logarithm terms truncates at one-loop order and fails to alter
the asymptotics already obtained;  {\it i.e.}
\begin{equation}
S_0 \left[g^2(\mu) \log (\mu\rho)\right] = 1 + 2b_3 g^2 (\mu) \log
(\mu\rho) \equiv w.
\end{equation}

For summations of subsequent subleading logarithms, one may rewrite the
RG-equation (6.4) as
\begin{equation}
0 = \sum_{k=0}^\infty g^{2k} \left[ \frac{dS_k}{du} + \sum_{\ell=0}^k
b_{3+2(k-\ell)} \left[ (2\ell - 2) S_\ell + 2u \frac{dS_\ell}{du}
\right]\right],
\end{equation}
where we have replaced $g^2(\mu) \log (\mu\rho)$ with the variable $u$.
Since the right-hand side of (6.9) is a power series in $g$ that must
vanish order-by-order, we find after a little algebra that $S_k$
satisfies a first-order inhomogeneous linear differential equation in
the variable $w = 1 + 2b_3 u$,
\begin{equation}
\frac{d S_k}{dw} + \frac{(k-1)}{w} S_k = - \frac{1}{2b_3 w} \left[
\sum_{\ell=0}^{k-1} b_{3+2(k-\ell)} \left[ 2(\ell-1) S_\ell + 2(w-1)
\frac{dS_\ell}{dw}\right]\right], \; \; \; (k \geq 1)
\end{equation}
with initial conditions following from the definition (6.3) $\left[ u =
g^2 (\mu) \log (\mu\rho)\right]$
\begin{equation}
\lim_{w \rightarrow 1} S_k = \lim_{u \rightarrow 0} S_k = T_{k,0}.
\end{equation}
The solution to eq.\ (6.10) for $S_k [w]$  with this initial condition is given by
\begin{eqnarray}
S_k [w] & = & T_{k,0} / w^{k-1}\nonumber\\
& - & \frac{1}{2b_3 w^{k-1}} \int_1^w dt \; \; t^{k-2} \left\{
\sum_{\ell=0}^{k-1} b_{3+2(k-\ell)} \left[ (2\ell-2) S_\ell [t] + 2(t-
1) S_\ell^{\prime} [t] \right] \right\}.
\end{eqnarray}
We see from eq.\ (6.8) that
\begin{equation}
S_1 = T_{1,0} + \frac{b_5}{b_3} \log |w|
\end{equation}
\begin{equation}
S_2 = \left( \frac{b_7}{b_3} - \frac{b_5^2}{b_3^2} \right) +
\frac{T_{2,0} - \left[ \frac{b_7}{b_3} - \frac{b_5^2}{b_3^2} \right] +
\frac{b_5^2}{b_3^2} \log |w|}{w}.
\end{equation}
Moreover, we find that the curly-bracketed contribution to the integrand
of (6.12) arising from eqs.\ (6.8), (6.13) and (6.14) is of the form $\{
... \}$ = constant $+ \; {\cal{O}}(1/t)$, which ensures that $S_3 [w]$ is
of the same ``constant $+ \; {\cal{O}}(1/w)$'' form as $S_2 [w]$.  Hence
the $\ell = 3$ contribution to the curly-bracketed expression to (6.12),
and by iteration all subsequent-$\ell$ contributions, will be of the
form $\{ ... \}$ = constant $+ \; {\cal{O}}(1/t)$, in which case we easily
see from eq.\ (6.12) that in the large-{\it w} limit
\begin{equation}
S_k [w] = c_k + {\cal{O}} (1/w), \; \; k \geq 2.
\end{equation}
Note that $c_k$ is a constant entirely obtained from $\beta$-function
coefficients, as in eq.\ (6.14), and does not depend on the coefficients
$T_{k,0}$ calculated from field theory.

If we substitute the results (6.8), (6.13) and (6.15) into the series
$S$ appearing in (6.1), we find that the $\rho$-dependence of the
integrand of (6.1) in the $\rho \rightarrow 0$ or $\rho \rightarrow
\infty$ limit is
\begin{eqnarray}
\rho^{-5+3n_f} \exp \left[ -\frac{8\pi^2}{g^2} \sum_{k=0}^\infty g^{2k}
S_k \right]\nonumber\\
        \begin{array}{c}{} \\ \longrightarrow\\ 
        _{| \log \rho | \rightarrow \infty}
        \end{array}
\rho^{\frac{7}{3} (1 + n_f) - {s.f.}} | \log \rho |^{-
8\pi^2 b_5 / b_3}.
\end{eqnarray}
The exponent of $\rho$, which is obtained in ref.\ [16], follows via eq.\
(6.6) entirely from the leading-log $T_{1,1} g^2 \log (\mu
\rho)$ contribution.  The exponent of $|\log \rho |$ follows from the
summation (6.13) of next-to-leading logarithms;  all subsequent summations (6.15)
of logarithms generate contributions independent of $\rho$ in the large-$\rho$
limit.  Thus, the result (6.16) is obtained via
summation of {\it all} logarithms contributing to the series $S$, and is
independent of any requirement that $g^2(\mu)$ be small.  Surprisingly,
we find only a very minimal amelioration of the infrared (large-$\rho$)
divergent behaviour characterising the integral (6.1).  Moreover, if the
scalar field (s.f.) contribution were sufficiently large to render the exponent
of $\rho$ in eq.\ (6.16) negative, the resulting ultraviolet divergence
at $\rho = 0$ would persist upon incorporating the summation of higher order logarithms.

\section*{Acknowledgements}
We are grateful for discussions with V. A. Miransky, and for research 
support from the Natural Sciences and Engineering Research Council of Canada. 
R.\ and D.\ MacKenzie provided helpful suggestions.

\newpage

\begin{figure}[htb]
\centering
\includegraphics[scale=0.6]{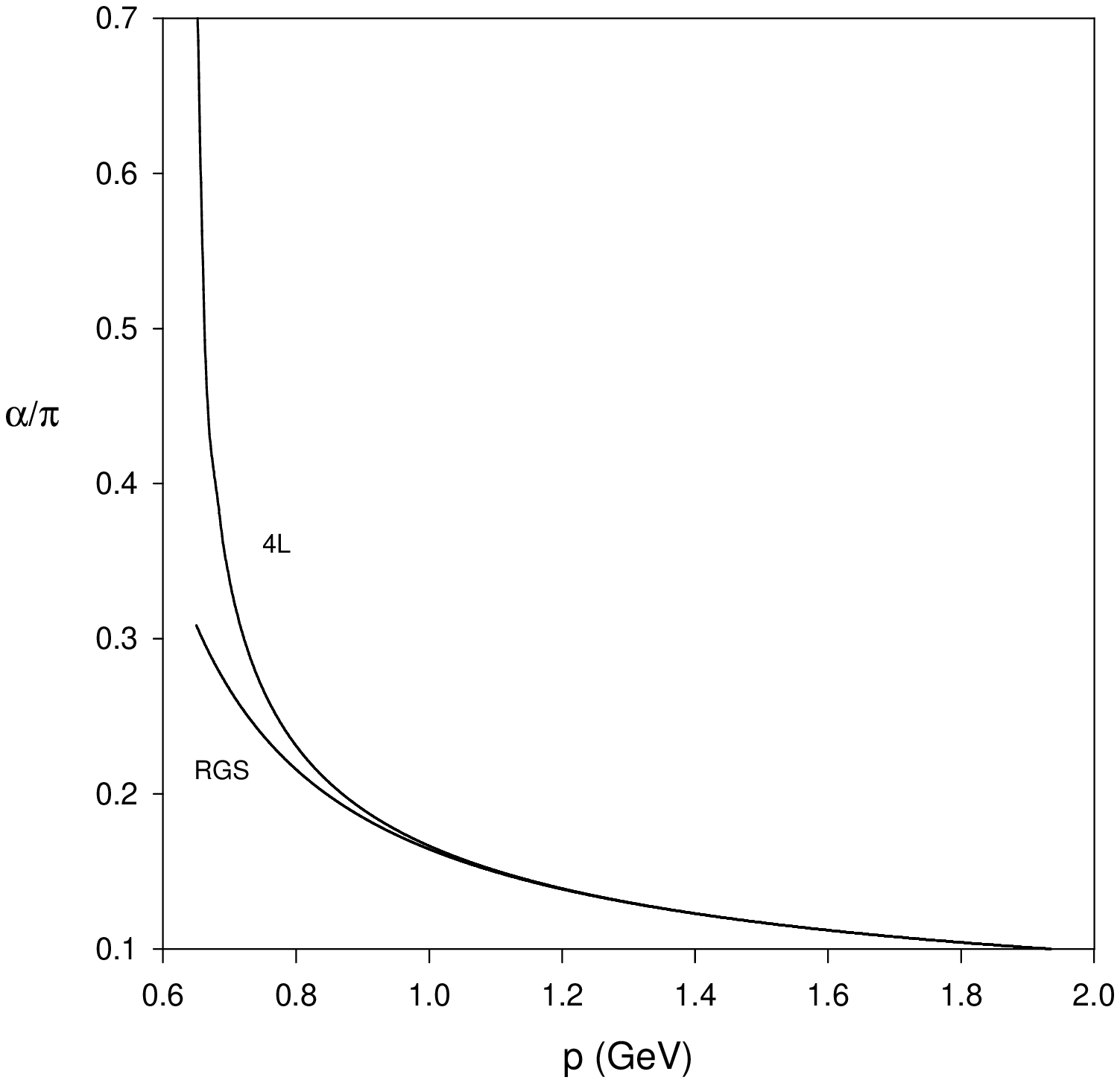}
\caption{Comparison of the true four-loop (4L) effective coupling constant, as obtained
via integration of the four-loop $\beta$-function, to the coupling constant obtained
in eq.\ (5.14) via summation of successively subleading logarithmic terms (RGS).}
\label{Figure 1}
\end{figure}

\begin{center}
\begin{figure}[htb]
\centering
\includegraphics[scale=0.6]{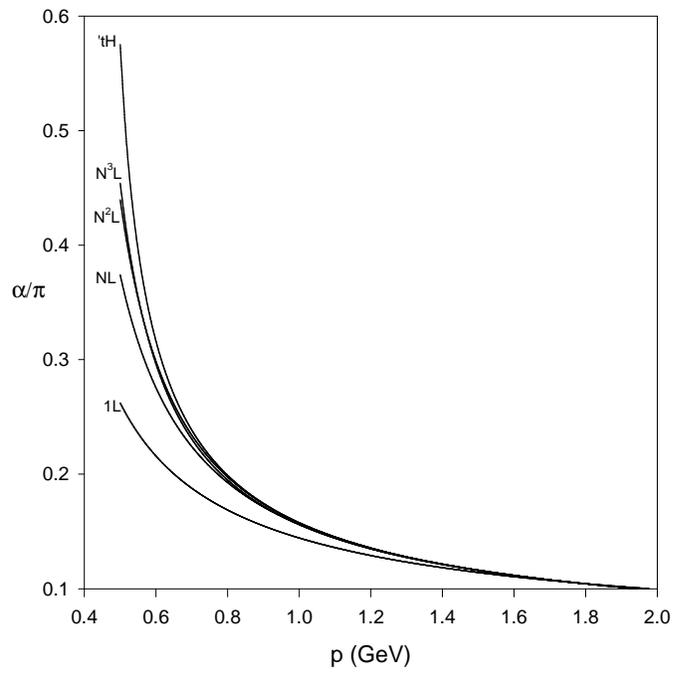}
\caption{Comparison of successive subleading approximations to the 't Hooft
scheme effective couplant, as discussed in the text, to the exact evolution
of the 't Hooft scheme ('tH) couplant.}
\label{Figure 2}
\end{figure}
\end{center}

\newpage

\section*{Appendix A:  ${\cal{O}} (\lambda^2)$-Equivalence of Two Expressions for $A(\lambda, L)$}

Two ``lowest-order''  RG-improved expressions for 
$A(\lambda, L)$ in the $\phi_6^3$ effective action (2.2) are

\renewcommand{\theequation}{A.\arabic{equation}}
\setcounter{equation}{0}

\begin{equation}
A(\lambda, L) = x^{D_2/B_3},
\end{equation}
\begin{equation}
A(\lambda, L) = \left[ \frac{K W \left[ \frac{2}{K} \exp \left( \frac{2x}{K} \right) \right]}{2}\right]^{D_2/B_3},
\end{equation}
where
\begin{equation}
x \equiv 1 - B_3 \lambda^2 L, \; \; K \equiv G_2 \lambda^2.
\end{equation}
[Eqs.\\ (A.1) and (A.2) are eqs.\ (2.28) and (3.15) of the text, respectively.]  In this appendix
we will demonstrate that eqs.\ (A.1) and (A.2) differ by at most ${\cal{O}}(\lambda^4)$.

We first note that both expressions are equal to unity when $L = \log \left( \frac{m^2}{\mu^2} \right) = 0$,
hence, when $x = 1$.  To see this for the second expression, we note that the Lambert W-function in eq.\ (A.2)
satisfies the constraint
\begin{equation}
W \left[ \frac{2}{K} e^{2/K} \right] \exp \left( W \left[ \frac{2}{K} e^{2/K} \right] \right) = \frac{2}{K} \exp \left( \frac{2}{K} \right),
\end{equation}
in which case $W \left[ \frac{2}{K} e^{2/K} \right] = 2/K$, and $A(\lambda, L)$, as defined by eq.\ (A.2), is
equal to 1 when $x = 1$.

Suppose we now perturb eq.\ (A.2) about $K = 0$.  To compare eqs.\ (A.2) and (A.1) with $\lambda$ small, we
utilise the following lowest order deviation:
\begin{eqnarray}
\frac{KW \left[ \frac{2}{K} \exp \left( 2x/K \right) \right]}{2} & = & \lim_{K \to 0}
\left( \frac{K W \left[\frac{2}{K} \exp [2x/K] \right]}{2} \right)\nonumber\\
& + & K \lim_{K \to 0} \left[ \frac{d}{dK} \left( \frac{K W \left[ \frac{2}{K} \exp \left( \frac{2x}{K} \right) \right]}{2} \right)\right]\nonumber\\
& + & {\cal{O}} (K^2).
\end{eqnarray}
The first limit on the right-hand side of eq.\ (A.5) is $x$, and the second limit is $-\log (x) / 2$,
as we shall show below.  Consequently, we have
\begin{eqnarray}
\frac{KW \left[ \frac{2}{K} \exp \left( 2x/K \right) \right]}{2} & = & x - \frac{K}{2} \log (x) + {\cal{O}} (K^2)\nonumber\\
& = & x + \frac{G_2 B_3}{2} \lambda^4 L + {\cal{O}} (\lambda^4),
\end{eqnarray}
as $K^2 = G_2^2 \lambda^4$ and $\log x = - B_3 \lambda^2 L + {\cal{O}}(\lambda^4)$.  
Upon substitution of eq.\ (A.6) into eq.\ (A.2), we find that eq.\ (A.2) deviates from eq.\ 
(A.1) at most by order $\lambda^4$, as stated at the end of Section 3.

To complete this demonstration, we evaluate explicitly the limits on the right-hand side of eq.\ (A.5).  Defining
$\eta = 2/K$, we find via L'H\^opital's rule that
\begin{eqnarray}
L_1 & \equiv & \lim_{K \to 0} \frac{KW \left[ \frac{2}{K} \exp (2x/K)\right]}{2} = \; \; \lim_{\eta \to \infty} \frac{W\left[ \eta e^{\eta x}\right]}{\eta}\nonumber\\
& = & \lim_{\eta \to \infty} \frac{W^{\prime} \left[\eta e^{\eta x} \right] \left( 1 + \eta x\right) e^{\eta x}}{1}.
\end{eqnarray}
One can easily show from the defining relation $W(\xi) \exp [W(\xi)] = \xi$ that $W^{\prime} (\xi) = 
W(\xi) / \left[ \xi(W(\xi) + 1) \right]$, in which case
\begin{eqnarray}
L_1 & = & \lim_{\eta \to \infty} \frac{W(\eta e^{\eta x})(1+\eta x)}{\eta[1+W(\eta e^{\eta x})]}\nonumber\\
& = & \left( \lim_{W \to \infty} \frac{W}{1+W} \right)x = x,
\end{eqnarray}
justifying the first term on the right-hand side of eq.\ (A.6).

To evaluate the second limit on the right-hand side of eq.\ (A.6), we first note that
\begin{equation}
\frac{d}{dK} \left( \frac{KW\left[\frac{2}{K} \exp \left( \frac{2x}{K} \right) \right]}{2} \right) = 
\frac{W\left[\frac{2}{K} \exp \left( \frac{2x}{K} \right) \right] \left[ KW \left[ \frac{2}{K} \exp \left( \frac{2x}{K} \right) \right] 
- 2x \right]}{2K \left[W\left[\frac{2}{K} \exp \left( \frac{2x}{K} \right) \right] + 1 \right]}
\end{equation}
Since $\lim_{W \to \infty} W / (W+1) = 1$, we find that
\begin{eqnarray}
L_2 & \equiv & \lim_{K \to 0} \frac{d}{dK} \left( \frac{KW \left[ \frac{2}{K} \exp \left( \frac{2x}{K}\right) \right]}{2} \right) = \; 
\lim_{K \to 0} \left( \frac{K W \left[ \frac{2}{K} \exp \left( \frac{2x}{K} \right) \right] - 2x}{2K} \right)\nonumber\\
& = & \lim_{\eta \to \infty}\frac{\left( W \left[ \eta e^{\eta x} \right] - \eta x \right)}{2}
\end{eqnarray}
We exponentiate eq.\ (A.10) and use the Lambert function relation $\exp \left[ W(\xi)\right] = \xi / W (\xi)$ to find that
\begin{eqnarray}
\exp (2L_2) & = & \lim_{\eta \to \infty} \exp \left[ W \left[ \eta e^{\eta x} \right] \right] e^{-\eta x}\nonumber\\
& = & \lim_{\eta \to \infty} \left( \frac{\eta}{W \left[ \eta e^{\eta x} \right]} \right) = \frac{1}{x}
\end{eqnarray}
The final step of eq.\ (A.11) follows from the evaluation of $L_1 = \lim_{\eta \to \infty} \frac{W\left[\eta e^{\eta x} \right]}{\eta} = x$, as 
obtained from eqs.\ (A.7) and (A.8).  Taking the natural logarithm of both sides of eq.\ (A.11), we find that
\begin{equation}
L_2 = -\frac{1}{2} \log (x),
\end{equation}
justifying the second term on the right-hand side of eq.\ (A.6).

\newpage


\renewcommand{\theequation}{B.\arabic{equation}}
\setcounter{equation}{0}

\section*{Appendix B:  Explicit Derivation of the   RG-Resummed Effective Couplant}

In this appendix we  sum leading and three subsequent subleading 
orders of logarithmic terms 
within the series expansion

\begin{equation}
x(p) = x(\mu) \sum_{n=0}^{\infty} \sum_{m=0}^{n} T_{n,m} x^n (\mu)
\log^m \left( \mu^2 / p^2 \right)
\end{equation}
for the running couplant, as defined by the differential equation
\begin{equation}
\mu^2 \frac{dx}{d\mu^2} (\mu) = \beta \left[ x(\mu) \right] ,
\end{equation}
with $x(p)$ as an initial condition that is necessarily independent of
the variable $\mu$.  For example, if $\beta = - \beta_0 x^2 (\mu)$, one can solve
eq.\ (B.2) directly to obtain the explicit one-loop (1L) result
\begin{equation}
x_{1L} (p) = \frac{x(\mu)}{1 - \beta_0 x(\mu) \log (\mu^2 / p^2)}.
\end{equation}
When one goes beyond one loop order, however, the solution to eq.\ (B.2)
is no longer explicit, but implicitly defined via the constraint
\begin{equation}
\log \left( \frac{p^2}{\mu^2} \right) - \int_{x(\mu)}^{x(p)}
\frac{ds}{\beta(s)} = 0.
\end{equation}
In this appendix, we will apply the RG-equation
\begin{equation}
\mu^2 \frac{d}{d\mu^2} x(p) = 0
\end{equation}
to the series (B.1) to obtain an explicit series solution for $x(p)$ that includes
the summation of leading and up to three subsequent subleading orders of
logarithms.

The RG-equation may be expressed in the following form:
\begin{equation}
\left( \frac{\partial}{\partial L} + \beta (x) \frac{\partial}{\partial x} \right) \sum_{n=0}^{\infty} \sum_{m=0}^{n} T_{n,m} x^{n+1} L^m = 0 ,
\end{equation}
where $x \equiv x(\mu)$ and $L \equiv \log (\mu^2 / p^2)$.  We note
from eq.\ (B.1) that when $L$ is equal to zero, $\mu^2$ is equal to $p^2$
and $x(\mu) = x(p)$.  This initial condition implies that the series
coefficients {\it not} involving logarithms satisfy the relations

\begin{equation}
T_{0,0} = 1 \; \; \; T_{k, 0} = 0 \; \; \; (k \neq 0)
\end{equation}

Consider first the one-loop case in which $\beta(x) = - \beta_0 x^2$.
One then finds that the aggregate coefficient of $x^p L^{p-k} \; \; (k
\geq 2)$ on the left hand side of eq.\ (B.6) must vanish:

\begin{equation}
(p-k+1) T_{p-1, p-k+1} - \beta_0 (p-1) T_{p-2, p-k} = 0
\end{equation}
Thus if $T_{k-2, 0} = 0$, as is the case for $k \geq 3$, then the
recursion relation (B.8) guarantees that all coefficients $T_{p+k-2, p}$
within the series will vanish when $k > 2$.  Consequently, the only
surviving coefficients of the series are those when $k = 2$, i.e. the
diagonal coefficients $T_{p,p}$, in which case we see from the series
(B.1) that to one-loop order
\begin{equation}
x(p) = x(\mu) \sum_{n=0}^{\infty} T_{n,n} x^n (\mu) \log^n (\mu^2 /
p^2).
\end{equation}
We also see from eqs.\ (B.7) and (B.8) that
\begin{equation}
T_{n,n} = \beta_0 T_{n-1, n-1} = \beta_0^n.
\end{equation}
Thus, eq.\ (B.9) is seen to be a geometric series whose explicit sum
recovers the result (B.3).

Consider now the full $\beta$-function (5.4) for which the first four
coefficients $\{ \beta_0, \beta_1, \beta_2, \beta_3 \}$ are known for
QCD applications in the $\overline{MS}$ scheme \cite{20}.  Upon substitution of
the $\beta$-function (5.4) into the left-hand side of the RG equation (B.6), we obtain
the following recursion relations from the vanishing of the aggregate
coefficients of $x^p L^{p-2}$, $x^p L^{p-3}$, $x^p L^{p-4}$, and $x^p
L^{p-5}$, respectively:
\begin{equation}
T_{p,p} = \beta_0 T_{p-1, p-1}
\end{equation}
\begin{equation}
0 = (p-2) T_{p-1, p-2} - \beta_0 (p-1) T_{p-2, p-3} - \beta_1 (p-2)
T_{p-3, p-3}
\end{equation}
\begin{equation}
0 = (p-3) T_{p-1,p-3} - \beta_0 (p-1) T_{p-2, p-4} - \beta_1 (p-2) T_{p-
3, p-4} - \beta_2 (p-3) T_{p-4, p-4}
\end{equation}
\begin{eqnarray}
0 & = & (p-4) T_{p-1, p-4} - \beta_0 (p-1) T_{p-2, p-5} - \beta_1 (p-2) T_{p-3, p-5}\nonumber\\
& - & \beta_2 (p-3) T_{p-4, p-5} - \beta_3 (p-4) T_{p-5, p-5}.
\end{eqnarray}
To make use of these recursion relations, we first re-organize the
series (B.1) into the following form $\left[ x \equiv x(\mu), \; \; L
\equiv \log (\mu^2 / p^2) \right]$
\begin{equation}
x(p) = \sum_{n=0}^\infty S_n (xL) x^{n+1}, \; \; \;  S_n (xL) = \sum_{k=n}^\infty T_{k, k-n} (xL)^{k-n}.
\end{equation}
Since $T_{0,0} = 1$, the recursion relation (B.11) implies that 
\begin{equation}
S_0 (xL) = \sum_{k=0}^\infty T_{k,k} (xL)^k = 1 / (1 - \beta_0 xL).
\end{equation}

To evaluate the subsequent summations $S_n (u)$, where $n \geq 1$ and $u
\equiv xL$, consider first the recursion relation (B.12). 
If we multiply this expression by $u^{p-3}$ and sum over $p$ from $p =
3$ to $\infty$, we obtain
\begin{eqnarray}
&& \sum_{p=3}^\infty (p-2) T_{p-1, p-2} u^{p-3} - \beta_0 \sum_{p=3}^\infty
(p-1) T_{p-2, p-3} u^{p-3}\nonumber\\
&& - \beta_1 \sum_{p=3}^\infty (p-2) T_{p-3, p-3} u^{p-3} = 0.
\end{eqnarray}
Using the definitions (B.15) for $S_0$ and $S_1$, we find that eq.\
(B.17) is the following first order linear differential equation for
$S_1 (u)$:
\begin{equation}
(1 - \beta_0 u) \frac{dS_1}{du} - 2 \beta_0 S_1 = \beta_1 \left[ u
\frac{dS_0}{du} + S_0\right],
\end{equation}
where $S_0 (u)$ is given by eq.\ (B.16) and where $S_1 (0) = T_{1,0} = 0$ by
eqs.\ (B.7) and (B.15).  The solution to this differential equation is
\begin{equation}
S_1 (u) = -\frac{\beta_1}{\beta_0} \frac{\log(1 - \beta_0 u)}{(1 -
\beta_0 u)^2}.
\end{equation}

Similarly, we can multiply the recursion relations (B.13) and (B.14)
by $u^{p-4}$ and $u^{p-5}$, respectively, and then by summing these
respective equations from $p = 4$ and $p = 5$ to infinity.  Using the
definitions (B.15) for $S_n (u)$, we then obtain the following
differential equations:
\begin{equation}
(1 - \beta_0 u) \frac{dS_2}{du} - 3\beta_0 S_2 = \beta_1 \left( u
\frac{dS_1}{du} + 2 S_1 \right) + \beta_2 \left( u \frac{dS_0}{du} + S_0
\right),
\end{equation}
\begin{equation}
(1 - \beta_0 u) \frac{d S_3}{du} - 4 \beta_0 S_3 = \beta_1 \left( u
\frac{dS_2}{du} + 3S_2 \right) + \beta_2 \left( u \frac{dS_1}{du} + 2S_1
\right) + \beta_3 \left( u \frac{dS_0}{du} + S_0 \right)
\end{equation}
with initial conditions (B.7):  $S_2 (0) = T_{2,0} = 0, \; \; S_3 (0) =
T_{3,0} = 0$.  The solution for $S_2 (u)$ is then found to be 
\begin{eqnarray}
S_2 (u) & = & \left( \frac{\beta_1^2}{\beta_0^2} - \frac{\beta_2}{\beta_0} \right) \left[ \left( 1 - \beta_0 u \right)^{-2} - \left( 1 - \beta_0 u \right)^{-3} \right]\nonumber\\
& - & \left( \frac{\beta_1}{\beta_0} \right)^2 \left( 1 - \beta_0 u \right)^{-3} \left[ \log \left(1 - \beta_0 u \right) - \log^2 \left( 1 - \beta_0 u \right) \right],
\end{eqnarray}
and the corresponding solution for $S_3$ is
\begin{eqnarray}
S_3 (u) & = & \left( -\frac{\beta_3}{2\beta_0} + \frac{\beta_1 \beta_2}{\beta_0^2} - \frac{\beta_1^3}{2\beta_0^3} \right) (1 - \beta_0 u)^{-2}\nonumber\\
& + & \left( \frac{\beta_1^3}{\beta_0^3} - \frac{\beta_1 \beta_2}{\beta_0^2} \right) (1 - \beta_0 u)^{-3}\nonumber\\
& + & \left( \frac{2\beta_1\beta_2}{\beta_0^2} - \frac{2\beta_1^3}{\beta_0^3} \right) \left( 1 - \beta_0 u \right)^{-3} \log \left( 1 - \beta_0 u \right)\nonumber\\
& + & \left( \frac{\beta_3}{2\beta_0} - \frac{\beta_1^3}{2\beta_0^3} \right) \left( 1 - \beta_0 u \right)^{-4}\nonumber\\
& + & \left( \frac{2 \beta_1^3}{\beta_0^3} - \frac{3\beta_1 \beta_2}{\beta_0^2} \right) \left( 1 - \beta_0 u \right)^{-4} \log \left( 1 - \beta_0 u \right)\nonumber\\
& + & \frac{\beta_1^3}{\beta_0^3} \left( 1 - \beta_0 u \right)^{-4} \left( \frac{5}{2} \log^2 \left( 1 - \beta_0 u \right) - \log^3 \left( 1 - \beta_0 u \right) \right).
\end{eqnarray}
Upon substitution of expressions (B.16), (B.17), (B.22) and (B.23) for $\{S_0$, $S_1$, 
$S_2$ \rm{and} $S_3\}$ into eq.\ (B.15), or alternatively, into eq.\
(5.9), we obtain eq.\ (5.14) for the RG invariant effective couplant
$x(p)$.

\end{document}